\documentclass[a4paper]{jpconf}
\usepackage{graphicx,url,amssymb,amsmath}

\newcommand{\pd}[3]{\frac{\partial^{#3} #1}{\partial {#2}^{#3}}} 
\newcommand{\td}[3]{\frac{d^{#3} #1}{d {#2}^{#3}}} 
\renewcommand{\v}[1]{\ensuremath{\mathbf{#1}}} 
\newcommand{\gv}[1]{\ensuremath{\mbox{\boldmath$ #1 $}}}

\begin{document}
\title{Dark matter gets DAMPE}

\author{Geoff Beck and Sergio Colafrancesco}

\address{School of Physics, University of the Witwatersrand, Private Bag 3, WITS-2050, Johannesburg, South Africa}

\ead{geoffrey.beck@wits.ac.za; sergio.colafrancesco@wits.ac.za}

\begin{abstract}
The DArk Matter Particle Explorer (DAMPE) recently reported an excess of electrons/positrons above expected background fluxes even when a double power-law background spectrum is assumed. Several dark matter models that involve TeV-scale leptophilic WIMPs have been suggested in the literature to account for this excess. All of these models are associated with the presence of a nearby dark matter clump/over-density.

In this work we set out to explore how current constraints from observational data impact the suggested parameter space for a dark matter explanation of the DAMPE excess, as well as make projections of the capacity of LOFAR and the up-coming SKA to observe indirect radio emissions from the nearby dark matter over-density. 

We show that LOFAR is incapable of probing the parameter space for DAMPE excess models, unless the dark matter clump is in the form of an ultra-compact mini halo. Fermi-LAT limits on dark matter annihilation are unable to probe these models in all cases. Limits derived from diffuse Coma cluster radio emission can probe a substantial portion of the parameter space and muon neutrino limits inferred from galactic centre gamma-ray fluxes heavily restrict muon coupling for the proposed WIMPs. The SKA is shown to able to fully probe the parameter space of all the studied models using indirect emissions from the local dark matter over-density.
\end{abstract}

\section{Introduction}
The DArk Matter Particle Explorer (DAMPE) has reported the direct detection of a break in cosmic ray electron spectrum at an energy $\sim 1$ TeV~\cite{dampe}. This corresponds to an excess in the electron/positron flux around $1.4$ TeV. Many recent works have proposed Dark Matter (DM) models to account for this excess via the annihilation of WIMPs and subsequent decay of a leptophilic mediator~\cite{dampedm1,dampedm2,dampeucmh}. A common element of all these models is that they contain leptophilic WIMPs of TeV mass and require the presence of an over-dense clump of DM within a radius $< 1$ kpc of the solar system in order for the DM models to also satisfy DM cosmological abundance constraints on the annihilation cross-section.


In this work we will explore the effects of current data on the parameter space for all the cases presented in \cite{dampedm1,dampedm2,dampeucmh}. We agree with the arguments in \cite{dampedm1,dampedm2,dampeucmh} that current gamma-ray limits from Fermi-LAT~\cite{Fermidwarves2015} don't have any substantial impact on constraining the proposed DAMPE excess DM models. However, we show that radio limits from the Coma cluster derived in \cite{gs2016}, and limits from muon neutrino fluxes inferred from gamma-ray fluxes from the galactic centre~\cite{gcneutrino} can make some impact on the allowed parameter space. We then explore the discovery potential for the necessary DM clumps with Low Frequency ARray (LOFAR) and the Square Kilometre Array (SKA) in radio. In so doing we show that only the SKA has the potential to completely rule out DAMPE models by hunting for radio emissions from the necessary local DM clump.

This paper is structured as follows: in section~\ref{sec:dampe} we discuss the candidate models in more detail, in sections~\ref{sec:dmann} and \ref{sec:emm} we detail the DM annihilation formalism and the subsequent radio emissions. In sections~\ref{sec:lim} and \ref{sec:sens} we provide information on the current constraints on DM we employ and the telescope sensitivities respectively. Finally, in sections~\ref{sec:res} and \ref{sec:conc} we display and discuss our results.

\section{Dark Matter Models for the DAMPE Excess}
\label{sec:dampe}

The DM models considered are heavy leptophilic WIMPs $\psi$ that couple to the Standard Model particles via a heavy mediator that is too large to allow for the decay of the WIMP~\cite{dampedm1,dampedm2}. Hence only annihilation will be considered here. We will consider the following ranges from the models listed above: $\psi$ couples to muons and electrons and spans a mass range around $1.4$ to $1.7$ TeV with cross-sections ranging from $3\times 10^{-26}$ to $5\times 10^{-24}$ cm$^3$ s$^{-1}$ in accordance with \cite{dampedm1}. The emissions stem from a DM clump of mass $10^6$ M$_{\odot}$ within a distance of $0.1$ kpc \cite{dampedm1} or a Ultra-Compact Mini-Halo (UCMH) of mass $\sim 3$ M$_{\odot}$ within a distance of $0.3$ kpc~\cite{dampeucmh}. For details of the UCMH formalism we refer the reader to \cite{ricotti2009,bringmann2012}. 

The second set of models considered has $\langle \sigma V \rangle  = 3 \times 10^{-26}$ cm$^3$ s$^{-1}$ with the electron only coupling ($e^+e^-$) and three lepton democratic coupling ($3l$) cases. For the $3l$ case we will work in the scenario of a DM clump situated at $0.3$ kpc with a mass of $2\times 10^8$ M$_{\odot}$. For the case of coupling to electrons only we use a halo with mass $8.0\times 10^{7}$ M$_{\odot}$ within a distance $0.3$ kpc. 

The distance and mass choices are representative of the models as a whole, as the distance and mass must co-vary to maintain the same flux in accounting for the excess observed by DAMPE. Non-UCMH clumps are considered to have NFW~\cite{nfw1996} density profiles with concentration parameters calculated according to \cite{prada2012}.

\section{Dark Matter Annihilation}
\label{sec:dmann}

The source function annihilation of WIMPs $\psi$ into final-state electrons with energy $E$ at halo position $r$ is given by
\begin{equation}
Q_i (r,E) = \frac{1}{2}\langle \sigma V\rangle \sum\limits_{f}^{} \td{N^f_e}{E}{} B_f \left(\frac{\rho_{\psi}(r)}{m_{\psi}}\right)^2 \; ,
\end{equation}
where $\langle \sigma V\rangle$ is the non-relativistic velocity-averaged annihilation cross-section at $0$ K, $B_f$ is the branching fraction for intermediate state $f$, $\td{N^f_e}{E}{}$ is the differential electron yield of the $f$ channel, and $\left(\frac{\rho_{\psi}(r)}{m_{\psi}}\right)^2$ is the WIMP pair number density.

The functions $\td{N^f_i}{E}{}$ will be sourced from \cite{ppdmcb1,ppdmcb2}. We will follow the standard practice of studying each annihilation channel $f$ independently, assuming $B_f = 1$ for each separate case (an exception is the $3l$ case where we weight each lepton channel equally). The studied channels will all be leptonic: $\tau$ leptons, muons, and electrons/positrons in accordance with \cite{dampedm1,dampedm2,dampeucmh}.


\section{Dark Matter Emissions}
\label{sec:emm}

The emission of interest in this work is synchrotron radiation produced by electrons sourced from DM annihilation and the subsequent decay of heavier products (if electrons are not directly produced). The average power of the synchrotron radiation at observed frequency $\nu$ emitted by an electron with energy $E$ in a magnetic field with amplitude $B$ is given by~\cite{longair1994}
\begin{equation}
P_{synch} (\nu,E,r,z) = \int_0^\pi d\theta \, \frac{\sin{\theta}^2}{2}2\pi \sqrt{3} r_e m_e c \nu_g F_{synch}\left(\frac{\kappa}{\sin{\theta}}\right) \; ,
\label{eq:power}
\end{equation}
where $m_e$ is the electron mass, $\nu_g = \frac{e B}{2\pi m_e c}$ is the non-relativistic gyro-frequency, $r_e = \frac{e^2}{m_e c^2}$ is the classical electron radius, and the quantities $\kappa$ and $F_{synch}$ are defined as
\begin{equation}
\kappa = \frac{2\nu (1+z)}{3\nu_g \gamma^2}\left[1 +\left(\frac{\gamma \nu_p}{\nu (1+z)}\right)^2\right]^{\frac{3}{2}} \; ,
\end{equation}
where $\nu_p \propto \sqrt{n_e}$ with $n_e$ being the density of the local thermal electron population, and
\begin{equation}
F_{synch}(x) = x \int_x^{\infty} dy \, K_{5/3}(y) \simeq 1.25 x^{\frac{1}{3}} \mbox{e}^{-x} \left(648 + x^2\right)^{\frac{1}{12}} \; .
\end{equation}

The magnetic field producing the synchrotron radiation will be taken to be that of the Milky-Way which hosts the DM clump producing the DAMPE excess. The field strength is thus taken to be the value in the solar neighbourhood $B = 1.5$ $\mu$G~\cite{mwbfield} and the thermal plasma density is given by $n_e = 0.03$ cm$^{-3}$~\cite{mwplasma}, these will be assumed constant over the volume of the DM clump or UCMH due to its small size. The Milky-Way magnetic field will be assumed to exhibit Kolmogorov-type turbulence with $1$ kpc coherence length.

The local emissivity for synchrotron emission can then be found as a function of the DM-produced electron and positron equilibrium distributions as well as the synchrotron power
\begin{equation}
j_{synch} (\nu,r,z) = \int_{m_e}^{M_\psi} dE \, \left(\td{n_{e^-}}{E}{} + \td{n_{e^+}}{E}{}\right) P_{synch} (\nu,E,r,z) \; ,
\label{eq:emm}
\end{equation}
where $\td{n_{e^-}}{E}{}$ is the equilibrium electron distribution from DM annihilation (see below).
The flux density spectrum within a radius $r$ is then written as
\begin{equation}
S_{i} (\nu,z) = \int_0^r d^3r^{\prime} \, \frac{j_{synch}(\nu,r^{\prime},z)}{4 \pi D_L^2} \; ,
\label{eq:flux}
\end{equation}
where $D_L$ is the luminosity distance to the halo.

In electron-dependent emissions there are two processes of importance, namely energy-loss and diffusion. Diffusion is typically only significant within small structures~\cite{Colafrancesco2007,gsp2015}, thus must be accounted for within the environment of the DM clumps of interest for the DAMPE excess.\\
The equilibrium electron distribution is found as a stationary solution to the equation
\begin{equation}
\begin{aligned}
\pd{}{t}{}\td{n_e}{E}{} = & \; \gv{\nabla} \left( D(E,\v{r})\gv{\nabla}\td{n_e}{E}{}\right) + \pd{}{E}{}\left( b(E,\v{r}) \td{n_e}{E}{}\right) + Q_e(E,\v{r}) \; ,
\end{aligned}
\end{equation}
where $D(E,\v{r})$ is the diffusion coefficient, $b(E,\v{r})$ is the energy loss function, and $Q_e(E,\v{r})$ is the electron source function from DM annihilation. In this case, we will work under the simplifying assumption that $D$ and $b$ lack a spatial dependence and thus we will include only average values for magnetic field and thermal electron densities. For details of the solution see \cite{Colafrancesco2007}.\\
We thus define the functions as follows~\cite{Colafrancesco1998}
\begin{equation}
D(E) = \frac{1}{3}c r_L (E) \frac{\overline{B}^2}{\int^{\infty}_{k_L} dk P(k)} \; ,
\end{equation}
where $\overline{B}$ is the average magnetic field, $r_L$ is the Larmour radius of a relativistic particle with energy $E$ and charge $e$ and $k_L = \frac{1}{r_L}$. This combined with the requirement that
\begin{equation}
\int^{\infty}_{k_0} dk P(k) = \overline{B}^2 \; ,
\end{equation}
where $k_0 = \frac{1}{d_0}$, with $d_0$ being the smallest scale on which the magnetic field is homogeneous, yields the final form
\begin{equation}
D(E) = D_0 d_0^{\frac{2}{3}} \left(\frac{\overline{B}}{1 \mu\mbox{G}}\right)^{-\frac{1}{3}} \left(\frac{E}{1 \mbox{GeV}}\right)^{\frac{1}{3}}  \; , \label{eq:diff}
\end{equation}
where $D_0 = 3.1\times 10^{28}$ cm$^2$ s$^{-1}$. We assume that $d_0 = 1$ kpc in order to make a conservative estimate of synchrotron emissions as the $d_0$ in the Milky-Way may be expected to be between $1$ pc and $1$ kpc~\cite{jansson2012a,jansson2012b}. 

The energy loss function is defined by
\begin{equation}
\begin{aligned}
b(E) = & b_{IC} E^2 (1+z)^4 + b_{sync} E^2 \overline{B}^2 \\&\; + b_{col} \overline{n} (1+z)^3 \left(1 + \frac{1}{75}\log\left(\frac{\gamma}{\overline{n} (1+z)^3}\right)\right) \\&\; + b_{brem} \overline{n} (1+z)^3 \left( \log\left(\frac{\gamma}{\overline{n} (1+z)^3 }\right) + 0.36 \right) \;,
\end{aligned}
\label{eq:loss}
\end{equation}
where $\overline{n}$ is the average thermal electron density in the UCMH and is given in cm$^{-3}$, while $b_{IC}$, $b_{synch}$, $b_{col}$, and $b_{brem}$ are the inverse Compton, synchrotron, Coulomb and bremsstrahlung energy loss factors, taken to be $0.25$, $0.0254$, $6.13$, and $1.51$ respectively in units of $10^{-16}$ GeV s$^{-1}$. Here $E$ is the energy in GeV and the B-field is in $\mu$G.

%
%

\section{Existing DM Limits}
\label{sec:lim}

We make use of the Fermi-LAT dwarf spheroidal galaxy limits on the annihilation cross-section~\cite{Fermidwarves2015}, as well as those derived from the diffuse radio emissions of the Coma cluster of galaxies~\cite{gs2016}. 

We also make use of the observed spectrum of the galactic centre in gamma-rays~\cite{fermigc2015} and infer from it a muon neutrino flux following \cite{gcneutrino}. This flux is then used to derive upper limits on the DM annihilation cross-section through comparison to predicted fluxes from the galactic centre using the same galactic halo parameters as used in \cite{fermigc2015}.

\section{Telescope Sensitivities}
\label{sec:sens}

We compare sensitivity projections for the SKA~\cite{ska2012} and LOFAR\footnote{\url{http://www.astron.nl/radio-observatory/astronomers/lofar-imaging-capabilities-sensitivity/sensitivity-lofar-array/sensiti}} to determine the smallest value of $\langle \sigma V \rangle$ that would be detectable at a confidence level of $5 \sigma$. 

In the case of the clumps with NFW profiles we must account for flux loss by these interferometers due to the minimum baseline effect (UCMHs are too compact to suffer from this). To this effect we use the fact that the Westerbork Synthesis Radio Telescope (WSRT) was found to capture only $0.16$ of the radio flux emitted by the nearby M33 object~\cite{viallefond}. Thus, we will make use of an ansatz that the flux captured by radio interferometry can be estimated as
\begin{equation}
f_{cap} = 0.16 \frac{\theta_{M33}}{\theta_{halo}} \frac{l_{WSRT}}{l_{obs}} \; ,
\end{equation}
where $\theta_{M33}$ and $\theta_{halo}$ are the angular sizes of M33 and the  angular radius containing $99\%$ of the flux from the DM clump (with a steeply peaked spatial profile reflecting that of an NFW density distribution squared), $l_{WSRT}$ and $l_{obs}$ are the minimum baseline lengths of WSRT and either the SKA or LOFAR. Using this calculation we find that $f_{cap} \sim 10^{-3}$ for clumps with NFW profiles at the distances given in Section~\ref{sec:dampe}.

\section{Results}
\label{sec:res}

In fig.~1 we display the results for the models from \cite{dampedm1,dampeucmh}, particularly upper-limits on $\langle \sigma V \rangle$ either compatible with known constraints on WIMPs or projections derived from potential non-observation of radio emissions from the local DM clump with a given instrument. We can see that the existing Fermi-LAT limits do not restrict the parameter space at all, as claimed by the authors of \cite{dampedm1}. However, some constraint is possible with the Coma limits from \cite{gs2016} and with the inferred neutrino flux form the galactic centre~\cite{fermigc2015,gcneutrino} for the case of the muon coupling only. LOFAR is unable to detect the extended clumps but is suggested to be able to probe the complete parameter space for the UCMH objects. The strongest possibility comes from the SKA, where it can find the radio emissions from annihilation the DM clumps for the entire parameter space.
 
\begin{figure}
	\centering
	\resizebox{0.6\hsize}{!}{\includegraphics{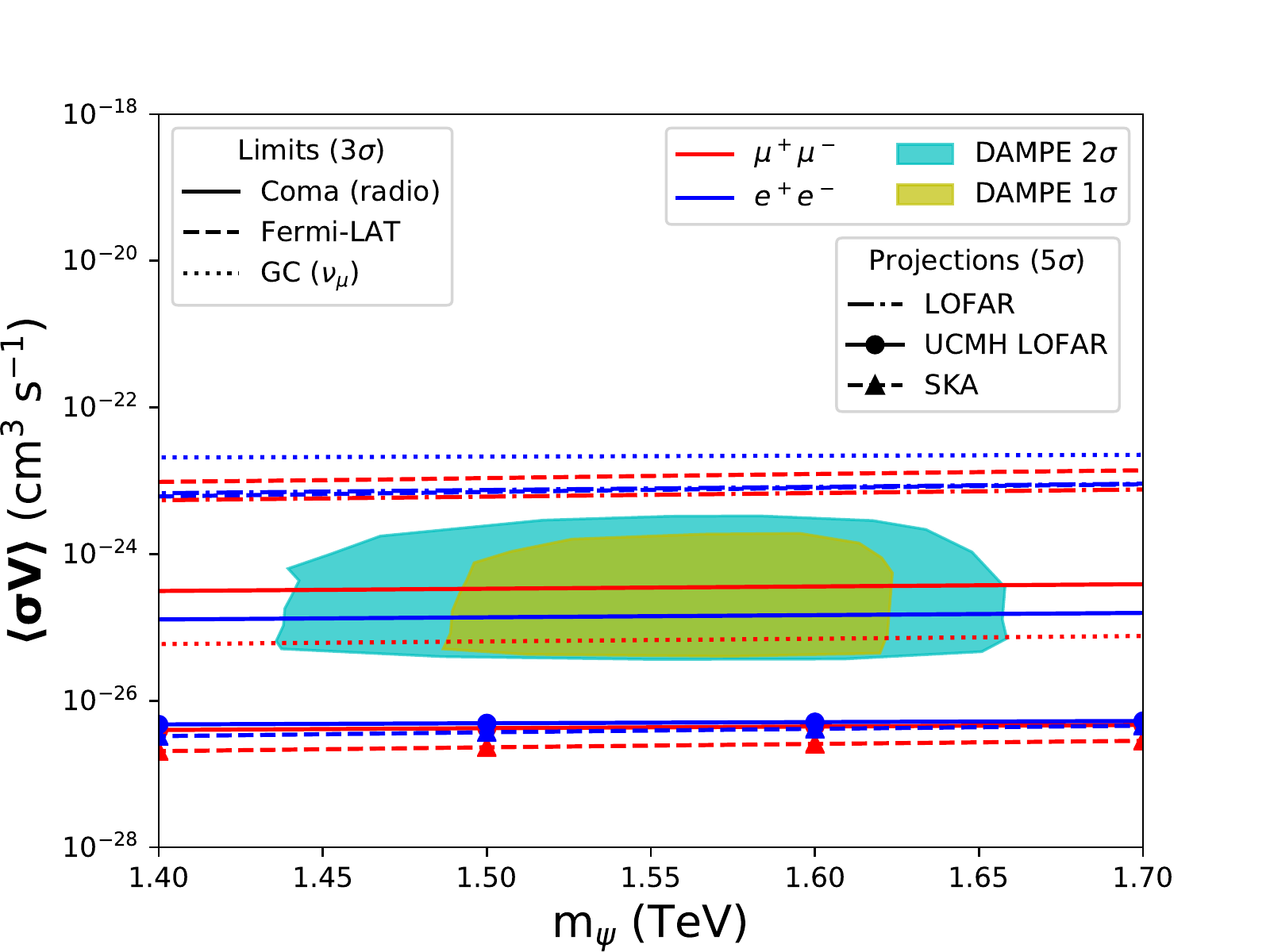}}
	\begin{minipage}[b]{14pc}\caption{Cross-section upper limits on DAMPE excess models from \cite{dampedm1,dampeucmh}. GC ($\nu_{\mu}$) refers to inferred galactic centre neutrino flux limits.}
	\end{minipage}
	\label{fig:dampe1}
\end{figure}

In fig.~2 we see the results for \cite{dampedm2}. In both the electrons only and the $3 l$ model only the SKA can reach below the thermal relic cross-section specified by the DM model and still detect the emissions of the DM clump. However, LOFAR gets close to the thermal relic cross-section for both cases and the Coma limits are similar for the $e^+e^-$ only case.
\begin{figure}
	\centering
	\resizebox{0.6\hsize}{!}{\includegraphics{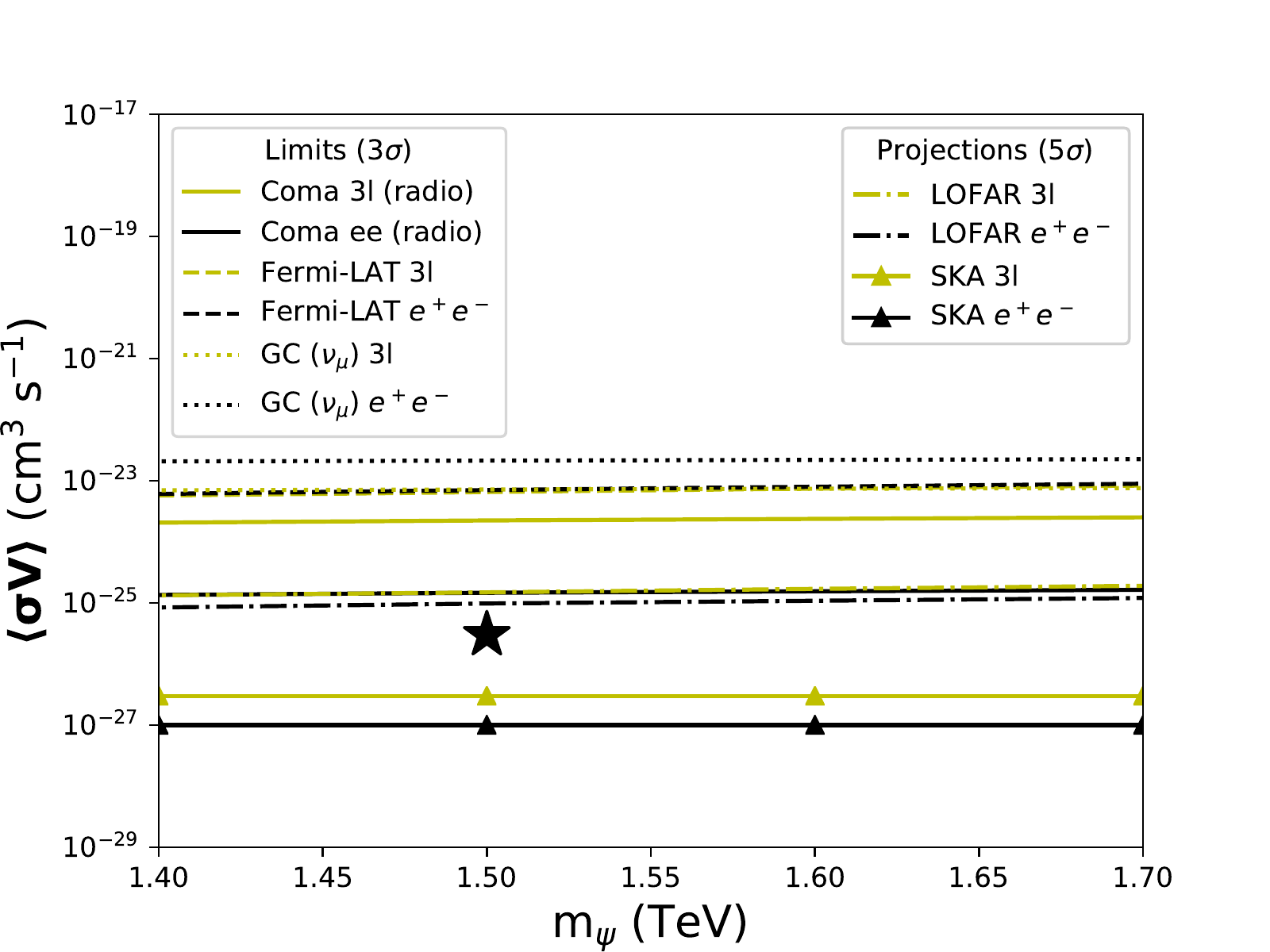}}
	\begin{minipage}[b]{14pc}\caption{Cross-section upper limits on DAMPE excess models from \cite{dampedm2}, the model parameter space is given by the star. GC ($\nu_{\mu}$) refers to inferred galactic centre neutrino flux limits. Note the Fermi-LAT lines lie nearly atop each other.}
	\end{minipage}
	\label{fig:dampe2}
\end{figure}

\section{Conclusions}
\label{sec:conc}
In conclusion we can see that new limits on the DM models conjectured to explain the DAMPE excess are possible with existing results from the Coma cluster and from inferred muon neutrino fluxes from the galactic centre. In particular, the inferred neutrino constraints provide little room for any muon coupling in the model of \cite{dampedm1}. However, Fermi-LAT is unable to observe gamma-ray emissions from the DM clumps that form part of these models. LOFAR can only detect the UCMH type clump from \cite{dampeucmh} due to flux loss over large extended sources. The SKA will be capable of probing the full parameter space for all the studied DAMPE excess models and would be able to rule them out in its phase 1 configuration by hunting for the radio emissions from the nearby DM clump that is necessary for all the studied models.

\ack 
This work is based on the research supported by the South African
Research Chairs Initiative of the Department of Science and Technology
and National Research Foundation of South Africa (Grant No 77948).

\section*{References}
\bibliographystyle{iopart-num}
\bibliography{dampe.bib}

\end{document}